\documentclass{aastex62}
\newcommand{\gtsim}{\raisebox{-1.0ex}{$\stackrel{\textstyle>}{\sim}$}}
\newcommand{\ltsim}{\raisebox{-1.0ex}{$\stackrel{\textstyle<}{\sim}$}}
\def\kms{km~s$^{-1}$}
\def\al{Alfv\'{e}n}

\def\hinode{{\sl Hinode}}

\def\p78{{\sl P78-1}}

\def\iris{{\sl IRIS}}

\def\feix{Fe~{\sc ix}}

\def\caii{Ca~{\sc ii}}

\def\cii{C~{\sc ii}}
\def\siiv{Si~{\sc iv}}
\def\mgii{Mg~{\sc ii}}
\def\halpha{H$\alpha$}

\def\al{Alfv\'{e}n}

\def\kms{km~s$^{-1}$}

\def\etal{et~al.}

\begin{document}
%

\title{Hi-C~2.1 Observations of Small-Scale Miniature-Filament-Eruption-Like Cool Ejections 
in Active Region Plage}

\correspondingauthor{Alphonse C.~Sterling}
\email{alphonse.sterling@nasa.gov}

\author{Alphonse C.~Sterling}
\affiliation{NASA/Marshall Space Flight Center, Huntsville, AL 35812, USA}

\author{Ronald L. Moore} 
\affiliation{NASA/Marshall Space Flight Center, Huntsville, AL 35812, USA}
\affiliation{Center for Space Plasma and Aeronomic Research, \\
University of Alabama in Huntsville, Huntsville, AL 35899, USA}

\author{Navdeep K. Panesar}
\affiliation{NASA/Marshall Space Flight Center, Huntsville, AL 35812, USA}
\affiliation{Bay Area Environmental Research Institute, NASA Research Park, Moffett Field, CA 94035, USA}
\affiliation{Lockheed Martin Solar and Astrophysics Laboratory, Palo~Alto, 
CA 94304, USA}

\author{Kevin P. Reardon}
\affiliation{National Solar Observatory,Boulder, CO 80303, USA}

\author{Momchil Molnar}
\affiliation{National Solar Observatory,Boulder, CO 80303, USA}
\affiliation{Department of Astrophysics and Planetary Sciences,
University of Colorado, Boulder, CO, 80303, USA}

\author{Laurel A. Rachmeler}
\affiliation{NASA/Marshall Space Flight Center, Huntsville, AL 35812, USA}

\author{Sabrina L. Savage}
\affiliation{NASA/Marshall Space Flight Center, Huntsville, AL 35812, USA}

\author{Amy R. Winebarger}
\affiliation{NASA/Marshall Space Flight Center, Huntsville, AL 35812, USA}

\begin{abstract}

We examine 172~\AA\ ultra-high-resolution images of a solar plage region from the Hi-C~2.1 (``Hi-C'') rocket flight of 2018 May~29.  
Over its five-minute flight, Hi-C resolves a plethora of small-scale dynamic features that appear near noise 
level in concurrent Solar Dynamics Observatory (SDO) Atmospheric Imaging Assembly (AIA) 171~\AA\ images.   
For ten selected events, comparisons with AIA images at other wavelengths and with the Interface Region Imaging Spectrograph 
(\iris) images indicate that these features are cool (compared to the corona) ejections.  Combining Hi-C 172~\AA, AIA~171~\AA, \iris\ 
1400~\AA, and \halpha, we see that these ten cool ejections are similar to the \halpha\ ``dynamic fibrils" 
and \caii\ ``anemone jets" found in earlier studies.
The front of some of our cool ejections are likely heated, showing emission in \iris\ 1400~\AA\@.  On average, these cool 
ejections have approximate widths: $3''.2\pm 2''.1$, (projected) maximum heights and velocities: $4''.3\pm 2''.5$ and 
$23\pm 6$~\kms, and lifetimes:
$6.5\pm 2.4$~min.  We consider whether these Hi-C features might result from eruptions of sub-minifilaments (smaller than
the minifilaments that erupt to produce coronal jets).  Comparisons with SDO's Helioseismic and Magnetic Imager (HMI) 
magnetograms do not show magnetic mixed-polarity neutral lines at these events' bases, as would be 
expected for true scaled-down 
versions of solar filaments/minifilaments. But the features' bases are all close to single-polarity
strong-flux-edge locations, suggesting possible local opposite-polarity flux unresolved by HMI\@. Or, it may be that
our Hi-C ejections instead operate via the shock-wave mechanism that is suggested to drive dynamic fibrils and
the so-called type~I spicules.

\end{abstract}


\section{Introduction}
\label{sec-introduction}

Observations of plage regions of the Sun in EUV at 171~\AA\ with the Solar Dynamics Observatory (SDO) Atmospheric 
and Imaging Assembly (AIA) hint at unresolved omnipresent dynamic activity of size scale of $\sim$1--few arcseconds,
varying on times scales near that channel's 12-s cadence.  Here we report on observations of resolved views of these
features from the 5-minute suborbital rocket flight of the High-Resolution Coronal Imager, version 2.1 (Hi-C~2.1; 
hereafter, ``Hi-C''), observing at 172~\AA\@.

Similar small-scale plage features have been observed at high resolution in \halpha\ \citep{berger_et99,depontieu_et99},
where they and similar features have been called various names, including ``dynamic fibrils" \citep[e.g.][]{depontieu_et07a} 
and ``type-1 spicules" \citep{depontieu_et07b}.  It is thought that many such \halpha\ features result from wave 
oscillations that evolve into shocks in the chromosphere \citep[e.g.,][]{hansteen_et06,depontieu_et07a}. They are also
similar to \caii\ chromospheric ``anemone jets" \citep{shibata_et07}, which are suggested to result from reconnection
between an emerging bipole (or any bipole) and ambient field \citep[e.g.,][]{shibata_et07,takasao_et13}.

Meanwhile, other 
recent studies have shown that solar coronal jets frequently result from eruption of small-scale filaments 
({\it minifilaments}), that are apparent in AIA EUV channels, including 171~\AA, of sizes ranging over 
$\sim$11$''$---$24''$ \citep{sterling_et15,panesar_et16}, and the production of jets from minifilament
eruptions has been modeled \citep{wyper_et17}.  Here, we will consider whether the events we resolve in Hi-C 
are cool ejections into the very low corona result from even-smaller-scale versions of such 
minifilament eruptions (which here we will call ``sub-minifilament eruptions'').


\section{Data Set}
\label{sec-data}

Hi-C flew on a sounding rocket on 2018 May~29, observing the Sun in 172~\AA\ \feix/{\sc x} emission 
for
about 5 minutes, with usable data from 18:56:26---19:01:43~UT \citep{rachmeler_et19}.  
Its field of view (FOV) covered active region NOAA AR 12712.  Here we examine a 
$40'' \times 40''$ 
subregion in the plage of the AR, containing only plage and a small sunspot (Figure~1(a---b)).  Hi-C
has $\sim$0$''.1$ pixels, and 4.4-s cadence \citep{rachmeler_et19}. Here we used movies with 
occasional jumps in cadence because we avoid images most strongly blurred by rocket jitter; 
the video accompanying Figure~1(c) shows that the resolution varied somewhat over the time 
period as a result.  Nonetheless,
we were able to follow the progress of several of the cool ejections, and we focus on
ten of them here.

In addition to Hi-C, we examined all seven AIA EUV channels of this subregion 
\citep{lemen_et12}.  As we will show, we could identify our ten Hi-C cool ejections in
AIA~171~\AA\@.  Otherwise however, the only AIA channels where we see  
counterparts to our Hi-C features are 304~\AA, 193~\AA, and weakly in 211~\AA, which 
are the AIA EUV channels 
with the coolest peak temperature response (respectively peaking at $\sim$5$\times 10^4$
and $1.5\times 10^6$~K), in addition to 171~\AA\ ($\sim$6$\times 10^5$~K).  (Thus, in the context of
this paper, by {\it cool ejections} we mean features that appear primarily in absorption in 172~\AA\
and 171~\AA\ images, with little or no signal in hotter channels.)  The AIA images have $0''.6$ pixels 
and 12-s cadence.  
We also examine slit-jaw images from the Interface Region Imaging Spectrograph (\iris) 
\citep{depontieu_et14} taken with $0''.17\times 2$ (double-binned) pixels at $\sim$13~s cadence 
in four filters: 1330~\AA\ (\cii, 30{,}000 K) and 1400~\AA (\siiv, 65{,}000~K), each with a 
40~\AA\ bandpass; and 2796~\AA\ (\mgii~k, 15{,}000~K) and 2831~\AA\ (\mgii~h/k wing, 
6000~K), each with a 4~\AA\ bandpass.  Additionally, we use line-of-sight magnetograms from the 
SDO Helioseismic and Magnetic Imager (HMI), which has $0''.5$~pixels and 45~s cadence 
\citep{scherrer_et12}.  See \S\ref{sec-discussion} for information on preliminary \halpha\ observations.

\section{Observations and Results}
\label{sec-observations}

Our selected FOV is full of fine-scale dynamic structure 
(see Figure~1(c) video).  In Figure~1(c), arrows show our ten selected features (Table~1), and 
boxes show their footpoints (i.e.\ where they appear to lift off the surface in Hi-C or AIA~171 videos).

We define the cool-ejection lifetimes (Table~1), as the time from the feature's first clear upward 
motion, until either it returned to the surface or faded.  (Strictly speaking, the motion we observe
is in the plane normal to our LOS\@.  Throughout this paper we assume that the actual motion is 
along the magnetic field extending from the 
surface into the corona, giving the features a basic up- and down-component to their motion,
such as for spicules.)  This duration frequently 
exceeded the 5-minute Hi-C flight.  Hence we 
supplemented our observations with movies made from AIA~171~\AA\ images, that extended 
several minutes before and after the Hi-C time window (Figure~1(d) and accompanying 
video).  For each event, we were able to identify the Hi-C features in AIA, albeit at substantially 
reduced resolution.  We used an early (pre-public-release) version of the Hi-C images that required
small manual shifts for alignment with the AIA images, and that had residual jitter resulting in 
jumps of order one-half arcsecond over the video's duration.  We estimate our final AIA-Hi-C 
alignment to be accurate to $\ltsim$1$''$ over our selected FOV\@.

\subsection{Hi-C 172~\AA\ Morphology}
\label{sec-Hi-C}

From extensive relatively recent investigations, prevailing views for the explanation of many chromospheric 
dynamic features in plage is that they are shock driven, or that they are driven by reconnection between 
emerging flux and ambient field; these ideas are discussed
at length in references of \S\ref{sec-introduction}, and we also consider these ideas further in 
\S\ref{sec-discussion}.  Here we consider whether the features we observe here, features that are prominent in Hi-C images, 
might be smaller cousins of the minifilament eruptions that drive coronal jets.  Because we now know that 
many jets are driven by eruptions of small-scale filaments, we consider it natural to entertain the possibility 
that the smaller jets we observe here might be driven by even smaller-scale eruptions; that is, we consider whether 
they might be sub-minifilament eruptions. In this Section we discuss the morphology of the features mainly in 
this context.  We first highlight two better-observed events: 3 and~9.

Event~3 begins near 18:58:27~UT (Figures~1(c,e) videos) and evolves into an upward-moving
prominent dark feature ($\sim$18:59:28~UT\@).  It has an obvious compact, circular shape, but
there may also be dark extensions of this feature to the southeast. Subsequently, 
the prominent dark feature rises with a (projected) velocity of $\sim$20~\kms, and starts to bend into a curved 
path near the end of the Hi-C video ($\sim$19:01:26~UT\@).  Its subsequent evolution is visible but harder to track 
in AIA~171~\AA, and by 19:04:57~UT it has faded and is no longer identifiable.

Event~9 has a somewhat different character.  It unambiguously begins as a bright, long and thin 
filament-like feature that rises upward from its western end ($\sim$18:56:56~UT; Figures~1(c,f) videos).  
A dark component 
appears adjacent to the brighter feature, eventually reaching (apparently) higher than and beyond 
the brighter feature ($\gtsim$18:59:28~UT\@).  That dark feature may eventually also follow a curved trajectory 
near the end of the Hi-C video ($\gtsim$19:00:29~UT\@). Continuing in AIA 171, the feature falls and 
fades away by 19:06:45~UT\@. 

It is plausible that event~3 is a more-compact version of event~9, with both being sub-minifilament eruptions.  
In support of this interpretation, event~3 is also similar to the compact roughly-circular ``blob" feature 
that produced a jet in \citet{adams_et14}.  In particular, Figure~6(c) of that paper shows the circular 
dark structure surrounded
by brightenings in 171~\AA\@.  Considering other evidence 
\citep[e.g.][]{shen_et12,sterling_et15,panesar_et16,panesar_et18a}, it is now almost certain that the
jet-producing blob of \citet{adams_et14} was an erupting minifilament.  Hence, even though the 
dark feature of event~3 is too amorphous to identify unambiguously as an erupting 
sub-minifilament, the additional jet studies suggest that it plausibly could be.  Thus,
the event-3 ejection may be a sub-minifilament eruption.  For event~9, its early brightenings 
could result from reconnections 
between the magnetic field enveloping the erupting filament-like feature with surrounding
field, resulting in a cocoon of brighter emission \citep[e.g.,][]{sterling_et11,li_et17}.

We next consider the remaining eight events.

Event-1's rise is visible in Hi-C, while its decay is only seen in AIA and hence is less clear.  This feature
rises in-sync with event~3, and their bases (boxes in Figure~1(c)) nearly overlap; thus the two events could 
be part of the same magnetic structure, although we cannot
identify a definitive connection.  Event~1 appears to be an upward-rising loop or (sub-mini)filament structure.
For example, at 19:00:29~UT, it appears as an absorbing feature $\sim$5$''$ long and $\sim$2$''$ tall.  
By 19:01:30~UT it has darkened, and the east side appears as a very short $\sim$3$''$-long jet structure.

Event~2 seems similar to event~9: it is an elongated emission feature, extending from the
arrow head to the edge of the FOV in the video accompanying Figure~1(c) at 18:57:53~UT\@.  
From 18:59:28~UT and until the end of the Hi-C movie, a portion of this feature 
ejects outward, north-northeastward, as a narrow ($\sim$1$''$) jet in emission.  This is similar to 
some narrow-spire coronal jets, whereby a minifilament can erupt upward into the 
base region of the jet, with jet material flowing out along 
a narrow spire (good examples are events 6 and 8 of \citeauthor{sterling_et15}~\citeyear{sterling_et15}).  
Therefore event~2 (and also likely event~9, and similar events) is a candidate for being a 
small-scale version of the minifilament eruptions that produce a coronal jet.

Event~4 is not fully resolved, but includes a weakly-emitting component moving westward over 
18:59:02---18:59:33~UT, followed by an upward rise of a curtain of
absorbing material, which remains elevated at the end of the Hi-C movie and is about
5$''$ wide and $2''.5$ high at its tallest.  This plausibly resembles a very weak failed/confined
(sub-mini)filament eruption.

Event~5 is a surge-like ejection that is larger than those so-far examined, and consequently 
(with insight from Hi-C) clearly identifiable in AIA~171.  Hi-C
only captures its return to the surface, but clearly shows a brighter portion at the 
feature's top.  From about 18:51:57~UT, AIA shows a bright feature leading the 
rise onset, similar to the initial brightenings in events~2 and~9.

Event~6, as also Event~5, is falling as the Hi-C movie starts (its apparent rise being clearly visible
in pre-Hi-C AIA images\@).  During the first 
30~s of Hi-C it seems to be part of Event~7, but by 18:57:22~UT, it has faded while
Event~7 has not; this, along with its earlier behavior over 18:53---18:55~UT in AIA~171,
makes this event appear to be distinct from event~7.  Again, there is a bright upper edge of the 
feature, clear in Hi-C (e.g.\ 18:56:34~UT).  From the start of the Hi-C movie until
18:57:00~UT, this event's absorbing portion appears to have a large horizontal extent ($\sim$5$''$), 
similar to event~1.

Event~7 also begins prior to Hi-C observations.  Over about 18:50---18:53~UT, AIA 
shows an emission feature moving upward in sync with the absorption material.  In this
case, Hi-C shows the absorption feature to have more of a long-narrow morphology 
(appearing as a small surge or spicule) than does event~6.

Event~8 roughly moves in sync with event~9, and their bases nearly overlap; so similar 
to the event~1 and~3 pair, these two features also may be part of the same magnetic structure 
(e.g., both may be in separate small, low-lying, adjacent bipoles, that reside beneath the same 
overlying magnetic enveloping field; eruption of one could trigger eruption of the second).  
Regarding them as separate events, event~8 is similar to event~1 in that it has a larger horizontal 
extent than vertical rise, and so as with event~1 it may be a confined eruption of a 
sub-minifilament structure.

Event~10 shows a clear bright front portion to an uprising absorption column, 
moving toward the northeast and forming a long and narrow
ejection.  AIA indicates that the feature reaches its maximum extent at about when
the Hi-C movie ends.

\subsection{AIA and \iris\ Morphology}
\label{sec-aia and iris}

We examined other AIA EUV channels for indications of the ten features.  Among these,
hints of the features were present in the ``cooler'' AIA channels, 193~\AA\ and 211~\AA,
while they were not visible in hotter channels such as 94~\AA\ and 131~\AA\@.  AIA~304~\AA\
shows well event~5, which in Hi-C is the largest and darkest of the absorbing features.
Also the emission components of events~2 and~9 are weakly discernible in 304 as they
erupt away from the surface.  None of the remaining seven features are unambiguously 
discernible from myriad surrounding and background not-fully-resolved dynamical motions
in the 304~\AA\ movies.

We manually aligned \iris\ images with features in AIA and
HMI\@.  Although \iris~1400 shows little obvious overlap with the bodies of our ten 
Hi-C events, close inspection reveals that tops of at least events~5, 9, and 10 are obvious 
in 1400 emission (Figure~3).  In some cases, such as
event~5, the 1400 emission is bright during the event's extension phase, but later 
(after 18:53:55~UT) the feature faded in 1400 even though it was still bright in Hi-C.  Matches of 
other events in 1400 were less obvious (perhaps due to a smaller projection angle).  
\iris~1330 shows similar morphology for events~5, 9, and~10, while corresponding features were difficult-to-detect 
or non-existent in the two remaining (cooler) \iris\ channels.

Overall, these findings support that some components of our features are heated to transition-region 
temperatures.

\subsection{Magnetic Setting}
\label{sec-magnetic}

We confirmed that the default HMI magnetogram alignment with AIA was accurate through 
comparisons with various features, including the sunspots in the full Hi-C FOV\@.  

Figure~2(a) shows the magnetogram with contours, and Figure~2(b) shows the same contours
on the Figure~1(c) Hi-C~172~\AA\ image.  Most apparent from the Hi-C overlay is that the plage appears
unipolar, the magnetogram showing only negative polarity.  It is also however immediately apparent that the negative-polarity 
strength is highly nonuniform, with stronger and weaker locations of negative
field throughout the plage region.  Therefore, while we see that all of our ten cool ejections originate from
a single-polarity location in HMI, we also see that all of them are within a couple arcseconds or so (about
the reliability of the overlay alignment) of an ``edge'' between stronger and weaker negative field.  We will
return to this point in the Discussion.

\section{Summary and Discussion} 
\label{sec-discussion}



A preliminary comparison with 
\halpha\ line-core intensity images of resolution comparable to Hi-C (but with highly varying seeing quality) obtained 
from the Dunn Solar Telescope at Sacramento Peak Solar Observatory, supplied by two 
of us (KPR and MM), shows that our Hi-C features do have \halpha\ components.  These \halpha\ features are 
nearly co-spatial with what we see with Hi-C, and display similar dynamics.
Thus our observed features plausibly make up some part of the population of small-scale dynamic fibrils observed
in \halpha\ and in EUV\@.  Also however, it is likely that they also coincide with some part of the population of
anemone jets observed in \caii\ \citep[cf.][]{morita_et10}.  Properties for the two features are very similar;
e.g., for chromospheric anemone jets, \citet{shibata_et07} give lengths of $\sim$3$''$---$7''$, and velocities 10---20~km/s,
which are very close to the Table~1 values for the Hi-C features ($\sim$2$''$---$7''$ and 10---20 km/s, respectively).  
The Hi-C features width ($1''$---$5''$) are larger than those of the \citet{shibata_et07} anemone jets ($0''.2$---$0''.4$),
which might be a consequence of the respective observed wavelengths.

Various studies suggest that many of the dynamic fibrils are driven
by acoustic waves that steepen into shocks in the chromosphere 
\citep[e.g.,][]{depontieu_et04,hansteen_et06,depontieu_et07a,depontieu_et07b}, resulting in oscillating
and repeated dynamic fibrils at close to the acoustic cutoff period, as in the rebound shocks
of \citet{hollweg82} (modified by radiation effects, e.g.\ \citeauthor{sterling_et90}~\citeyear{sterling_et90}).  
When the magnetic flux tube on which the field-aligned oscillations occur is sufficiently
inclined to the vertical, observations and numerical simulations support that the local acoustic cutoff 
period for acoustic waves along the flux tube is long enough to permit enhanced excitement
of waves (and hence shocks, and dynamic fibrils) along the tube at the photospheric $p$-mode oscillation period
\citep{suematsu90,depontieu_et07b,martinez-sykora_et09,heggland_et11}.

Even if many dynamic fibrils are driven by shocks, some other dynamic
features originating in the chromosphere could be due to a different mechanism, such as magnetic reconnection;
this might not be surprising, but would require a mixture of magnetic polarities in plage. There is substantial
work indicating that many spicules {\it outside of plages} are driven by a reconnection mechanism; these 
features have been called ``type~II spicules'' \citep{depontieu_et07b,pereira_et12,martinez-sykora_et17}.  
Some portion of at least those non-plage small-scale
dynamic events might fall on a continuum of reconnection-driven spicule-sized
features, macrospicules, and various surges and jets 
\citep[e.g.,][]{canfield_et96,chae_et98,hansteen_et06,depontieu_et07a,rouppe_et07,shibata_et07,depontieu_et07a,sterling_et16}.  

On the other hand, various different studies suggest that the anemone jets result from emerging flux reconnecting with ambient
field \citep[e.g.,][]{shibata_et07,nishizuka_et08,morita_et10,nishizuka_et11,singh_et12}.  These jets have an ``inverted-Y" 
shape at their base, which is consistent with the emerging flux model for jets (or any jet model based on reconnection
between a bipole and ambient field) \citep{shibata_et94,yokoyama_et95,moreno_et08}.
Because our observed Hi-C features resemble anemone jets, it is possible that undetected emerging flux is the driver
of our Hi-C events.

Regarding coronal jets, especially in coronal holes and quiet Sun, there exist strong observational evidence
that often the jets result from minifilament/flux rope eruptions, where the minifilament/flux ropes are built 
and triggered to erupt by magnetic flux cancelation \citep[e.g.][]{shen_et12,young_et14,sterling_et15,panesar_et16,panesar_et18a,mcglasson_et19}. In some 
cases however the minifilament/flux rope might be triggered to erupt by a different process \citep[e.g.,][]{kumar_et19}.
And we cannot rule out the possibility that some jets might result from the flux-emergence process in the absence
of clear flux cancelation.


Because many coronal jets result from flux-cancelation-built-and-triggered minifilament eruptions, in this work 
we are asking whether the smaller-than-jet cool ejections that we observe {\it in plage}, being prominent 
in Hi-C 172 images, might 
be driven by eruptions of smaller-scale
filament-like features that are smaller than the minifilaments that erupt to drive coronal jets; we have focused
on this viewpoint in this paper.
Perhaps the most glaring lack of evidence for the erupting (mini)filament idea however is that 
(mini)filaments always occur at the inversion line of mixed-polarity locations \citep[e.g., for 
jet minifilaments see][]{panesar_et16,panesar_et18a}.  Another obvious lack is that there is no 
definitive bright point at the base of our cool ejections corresponding to the jet bright point (JBP) often 
appearing at jet bases \citep{shibata_et92,sterling_et15}. 


Nonetheless, because of the non-uniform nature of the (seemingly) unipolar field of the plage region 
(Figure~2), it is possible that weak, small-scale opposite-polarity
intrusions exist in the plage.  In Figure~2, these would take the form of positive-polarity 
elements too small/weak to be detected with HMI, but that could result in localities in the
HMI magnetogram having weaker negative-field strength than in the surrounding plage.  Since 
our ten cool ejections are all rooted within a short distance of a field-intensity
edge (Figure~2), it is possible that small-scale positive-polarity elements 
might exist in those regions 
and give rise to the features we observe, as that flux cancels with surrounding dominant 
(negative) polarity.  Similarly, brightenings corresponding to JBPs
could exist but be very weak and subtle or hidden, as the magnetic reconnection that would be expected
to be the source of those brightenings would occur in a denser region of the atmosphere
than in the coronal jet case, with radiative losses from the dense material robbing much of the 
energy that would otherwise appear as a brightening (\citeauthor{sterling_et16}~\citeyear{sterling_et16} 
discuss a similar situation regarding weak brightenings, 
and \citeauthor{panesar_et18b}~\citeyear{panesar_et18b} discuss similar small-scale flux patches and brightenings).

We briefly consider our events in terms of the still-speculative suggestion of \citet{sterling_et16} that eruptions of
filament(-like) sheared/twisted magnetic-field structures of size scales of filaments,  minifilaments, and so-far-unseen
microfilaments might respectively drive CMEs, coronal jets, and many spicules.  This is consistent with suggestions of
\citet{moore_et77}, \citet{moore90}, and \citet{shibata99}.   Plotting  the number on the Sun at a given time of each
category of erupting feature against its size shows an approximate power law \citep[Figure~2 of][]{sterling_et16}.  We can add
a point to this plot with our cool ejections: simply extrapolating ten events over the $40''\times 40''$ Figure~1(c) FOV to the
entire  solar surface implies an upper bound of $\sim$70{,}000 cool events ($\pm$50\% as an uncertainty estimate) occurring on 
the Sun at any point in time.  These values, with  the average width in Table~1 of $\sim$2300$\pm$1500~km, gives the red point
in Figure~4.  Our new point is offset above the original power-law curve, but not totally inconsistent with it considering the
crudeness of the estimates.  The offset might also suggest a higher proportion of such events in plages, compared to the
non-plage spicule and jet estimates of \citet{sterling_et16}.  We add two caveats: first, because usually  only a small
fraction of the Sun, $\sim$10\%, is covered by plage and magnetic network flux, our estimate of the number events on the Sun
could be a factor of ten too high, and thus we extend a dashed line on the plot down to that reduced value.  Second, the number
of spicules might be higher than shown here ($\sim$2$\times 10^7$, according to \citeauthor{judge_et10}~\citeyear{judge_et10}).

We do emphasize however that it has not been established observationally how spicules 
are driven.  In addition to the ideas cited above, other 
possibilities discussed include \al\ waves \citep[e.g.,][]{hollweg_et82,kudoh_et99,matsumoto_et10,cranmer_et15,iijima_et17,martinez-sykora_et17},
or that spicules are sheet-like magnetic structures \citep{judge_et12}.
Moreover, the question of whether there are two types of spicules, type~I and type~II as defined in
\citet{depontieu_et07b}, is still unresolved; see \citet{zhang_et12} and \citet{pereira_et12} for opposing 
viewpoints; also see \citet{anan_et10}.  Also, the relationship between the pre-\hinode-era ``classical'' spicules of, e.g., 
\citet{beckers72}, and these possible spicules types is still being sorted out \citep{sterling_et10,pereira_et13}.
More recent spicule reviews than \citet{beckers72} include \citet{sterling00} and \citet{tsiropoula_et12}.

We find that our Hi-C features show strong similarities to the dynamic fibrils referred to as 
``type-I spicules'' \citep{depontieu_et04,depontieu_et07b}; the average lifetimes (6.4 min, from 
Hi-C and AIA combined observations) and 
upward velocities (23~\kms) from of our features from Table~1 match well with the type~I parameters 
(3--7 min, and $< 40$~\kms).  The shocks suggested as being the drivers of dynamic fibrils are an
alternative to our Hi-C events being 
small versions of coronal jets.   The jets often result from magnetic-flux cancelations forming minifilaments 
\citep{panesar_et17} that are triggered to erupt by further cancelation 
\citep{panesar_et16,panesar_et18a,sterling_et17}, producing the jets \citep{sterling_et15}.
If however there are small-scale mixed-polarity elements embedded inside of the
seemingly-unipolar plage field, then another possibility is that the sloshing from
granular motions and photospheric $p$-mode oscillations could drive cancelation among small-scale elements, 
generating the Hi-C features (and perhaps type~I spicules) in a manner similar to how coronal jets
are produced \citep[compare][]{morton_et12}.  

More generally, because we do not observe clear evidence for mixed polarity, converging flux, or emerging
flux, at this time we cannot rule out any of a variety of other mechanisms for driving the small-scale Hi-C 
features, including various ideas for jets \citep[e.g.,][]{shibata_et86,yokoyama_et95,pariat_et09}, or any of the various
models for spicules.  Alternatively, the ejections could be a kind of plasmoid (or flux rope in 3D space) \citep[e.g.][]{bhatt_et09,ji_et11}.

Long-term detailed studies (preferably satellite based) with a high-resolution EUV instrument similar to Hi-C, 
combined with high-resolution chromospheric images and photospheric/chromospheric magnetograms, e.g.\ from 
CRISP, CHROMIS, BBSO, and the upcoming DKIST facility, and with state-of-the-art numerical simulations, 
will help determine whether the features we observe here are due to chromospheric shocks, jet-like minifilament 
eruptions, or some other mechanism.  A direct near-future test will be with the high-resolution
magnetograms from DKIST, which should tell whether small-scale mixed-polarity elements exist within strong
plage, and whether those elements undergo cancelation at locations near where cool ejections originate.

\acknowledgments

We thank B. De~Pontieu for very fruitful discussions, and an anonymous referee for a very detailed 
and helpful report.  A.C.S. and R.L.M. received funding from the 
Heliophysics Division of NASA's
Science Mission Directorate  through the Heliophysics Guest Investigators (HGI) Program, and 
the \hinode\ Project. N.K.P's is supported by NASA SDO/AIA funding (NNG04EA00C); and
previously by the NASA 
Postdoctoral Program at NASA/MSFC, administered by Universities Space Research 
Association under contract with NASA\@. We acknowledge the Hi-C~2.1 instrument team 
for making the second re-flight data available under the NASA Heliophysics Technology 
and Instrument Development for Science (HTIDS) Low Cost Access to Space (LCAS) 
program.  MSFC/NASA led the mission with partners including the Smithsonian 
Astrophysical Observatory, the University of Central Lancashire, and Lockheed 
Martin Solar and Astrophysics Laboratory. \iris\ is a NASA small explorer mission developed 
and operated by LMSAL with mission operations executed at NASA Ames Research center and 
major contributions to downlink communications funded by ESA and the Norwegian Space Centre.

\clearpage

\begin{deluxetable}{ccccccccc}
\tabletypesize{\footnotesize}
\tablecaption{Properties of Hi-C Small-Scale Cool Ejections \label{tab:table1}}
\tablehead{
\colhead{Event} & \colhead{x,y location\tablenotemark{a}} & \colhead{start\tablenotemark{b}} & \colhead{end\tablenotemark{b}} & 
\colhead{lifetime (min)\tablenotemark{c}} & \colhead{Type\tablenotemark{d}} &  \colhead{Width (arcsec)\tablenotemark{e}} 
& \colhead{Height (arcsec)\tablenotemark{e}}& \colhead{rise vel (km/s)}
}
\startdata
1  & -79, 249 &  18:59:54 & 19:05:21 & 5.5$\pm 0.5$ &  abs smf & 6.5 & 2.0 & $26\pm 5$\tablenotemark{g} \\ 
2  & -97, 254 &  18:56:15  & 19:02:33  & 6.3$\pm 1.0$ & emit smf & 7.0 & 7.5 & $27\pm 5$\tablenotemark{g} \\
3  & -78, 253 &  18:58:21  & 19:04:57 & 6.6$\pm 0.5$ & surge-like & 2.0 & 6.0 & $20\pm 4$\tablenotemark{g} \\
4  & -69, 252 &  18:58:45  & 19:03:45 & 5.0$\pm 0.5$ & surge-like & 2.0 & 2.0 & $26\pm 5$\tablenotemark{g}\\
5\tablenotemark{f}  & -94, 262 &  18:52:15  & 19:02:33  & 10.3$\pm 0.5$ & surge-like & 3.0 & 9.0 & $18\pm 4$\tablenotemark{h} \\
6\tablenotemark{f}  & -86, 259 &  18:53:45  & 18:57:57  & 4.2$\pm 1.0$ & abs smf & 1.5(?) & 2.5 & $24\pm 12$\tablenotemark{i} \\
7\tablenotemark{f}  & -82, 258 &  18:53:33  & 18:58:27  & 4.9$\pm 1.0$ & surge-like(?) & 1.2 & 3.0 & $30\pm 15$\tablenotemark{i} \\
8  & -77, 260 &  18:57:22  & 19:02:45  & 5.4$\pm 1.0$ & abs smf & 3.3 & 2.0 & $18\pm 4$\tablenotemark{g}  \\
9  & -75, 264 &  18:55:21  & 19:06:45  & 11.4$\pm 1.0$ & emit smf &3.5 & 4.0 & $12\pm 3$\tablenotemark{g} \\
10  & -68, 263 &  18:58:58  & 19:04:21  & 5.4$\pm 1.0$ & surge-like & 1.5 & 5.2 & $29\pm 6$\tablenotemark{g}  \\ 
\hline
Averages\tablenotemark{j}  & --- &  ---  & ---  & 6.5$\pm 2.4$ & ---  & 3.2$\pm 2.1$ & 4.3$\pm 2.5$ & 23$\pm 6$  \\ 
\enddata
\tablenotetext{a}{Based on time of frame in Figure~1(c).}
\tablenotetext{b}{From Hi-C when in its observation window, and otherwise from AIA~171.  Estimated uncertainties generally $\sim$12--24~s.}
\tablenotetext{c}{Estimated uncertainties $\ltsim$1~m.}
\tablenotetext{d}{Description of dominant characteristic morphology.  Key: smf=sub-minifilament, abs=absorption, emit=emission.}
\tablenotetext{e}{Measured projected against disk.  Uncertainties generally $\leq 0.''5$.}
\tablenotetext{f}{Rise phase occurred before Hi-C observation window.}
\tablenotetext{g}{Rise velocities estimated to be accurate to $\sim$20\% when measured with Hi-C.}
\tablenotetext{h}{Rise velocity estimated from AIA, but event's large size reduces uncertainty ($\sim$20\%).}
\tablenotetext{i}{Rise velocity estimated from AIA, leading to relatively large uncertainty ($\sim$50\%).}
\tablenotetext{j}{Averages calculated assuming equal weights for all measurements.}

\end{deluxetable}
\clearpage


\figcaption{(a) Hi-C~2.1 (=``Hi-C'') 172~\AA\ image, showing subregions of the full FOV\@.  The black box is 
the $40''\times 40''$ subregion of panels (b), (c), and (d); and red and blue boxes are the respective $20''\times 20''$ 
subregions of panels (e) and (f).  (b)  HMI intensity 
image, showing a small sunspot and several pores in the FOV\@.  (c) Hi-C~172~\AA\ image in the boxed region
of (a).  Arrows and labels indicate cool ejection events of Table~1, with each event occurring (approximately) within the
accompanying green box. Black boxes indicate the apparent solar-surface footpoint 
locations for the ejection pointed to by the nearest arrowhead.  (d) AIA~171~\AA\ image, with boxes at the same
locations as those in (c).  Panels (e) and (f) are close-ups (red/blue boxes of (a)), with arrows pointing 
to (e) events 1 and 3, and (f) events 8 and 9. 
North is upward and west is to the right in all solar images of this publication.  
Panels~(c), (d), (e), and (f) are available as online animations.
Animations (c) and (d) run twice, once with arrows/boxes and once without. \label{fig1} }

\figcaption{(a) HMI magnetogram of the region shown in Figures~1(b,c,d), with black (white) indicating negative 
(positive) fields.  Overlaid green contours are negative fields at levels of -50 (thin contours), -100 (medium), and -750 
(thickest)~G\@.  Red contours show positive fields of corresponding levels, with only +50~G levels appearing 
in this image. Boxes are as in Figure~1(c). (b) Contours of (a) overlaid onto the Hi-C~172~\AA\ image of Figure~1(c). 
Both panels 2(a) and 2(b) are available as online animations. \label{fig2} }

\figcaption{\iris\ SJI 1400~\AA\ images of the Hi-C region of Fig.~1, with black and green arrows and boxes the same as 
in Fig.~1c. White arrows in (a) and (b) show evolution of a UV feature corresponding to event~5; similarly, in (c) and (d)
(with zoomed-in FOV) yellow and turquoise arrows show evolution of UV features corresponding to events 9 and 10.  An animation is
available. \label{fig3} }

\figcaption{Number of eruptive events on Sun at a given time as a function of 
erupting-(mini)filament size, for (left to right) spicules, the Hi-C events of this paper (open red circle), coronal jets,
and large-scale erupting filaments.  The values for the spicules and the Hi-C features are speculative, since small-scale filaments
have not been observed in such jets. The three black circles and the line (which is the best fit to those
three points) are the same as in \citet{sterling_et16}.  The bottom of the dashed red line is where the Hi-C point
would be if we reduced the extrapolated number on the Sun by a factor of ten (consistent with the number of plage 
cool ejections being overestimated by a factor of ten). \label{fig4} }

\clearpage

\begin{figure}
\epsscale{1.0}
\plotone{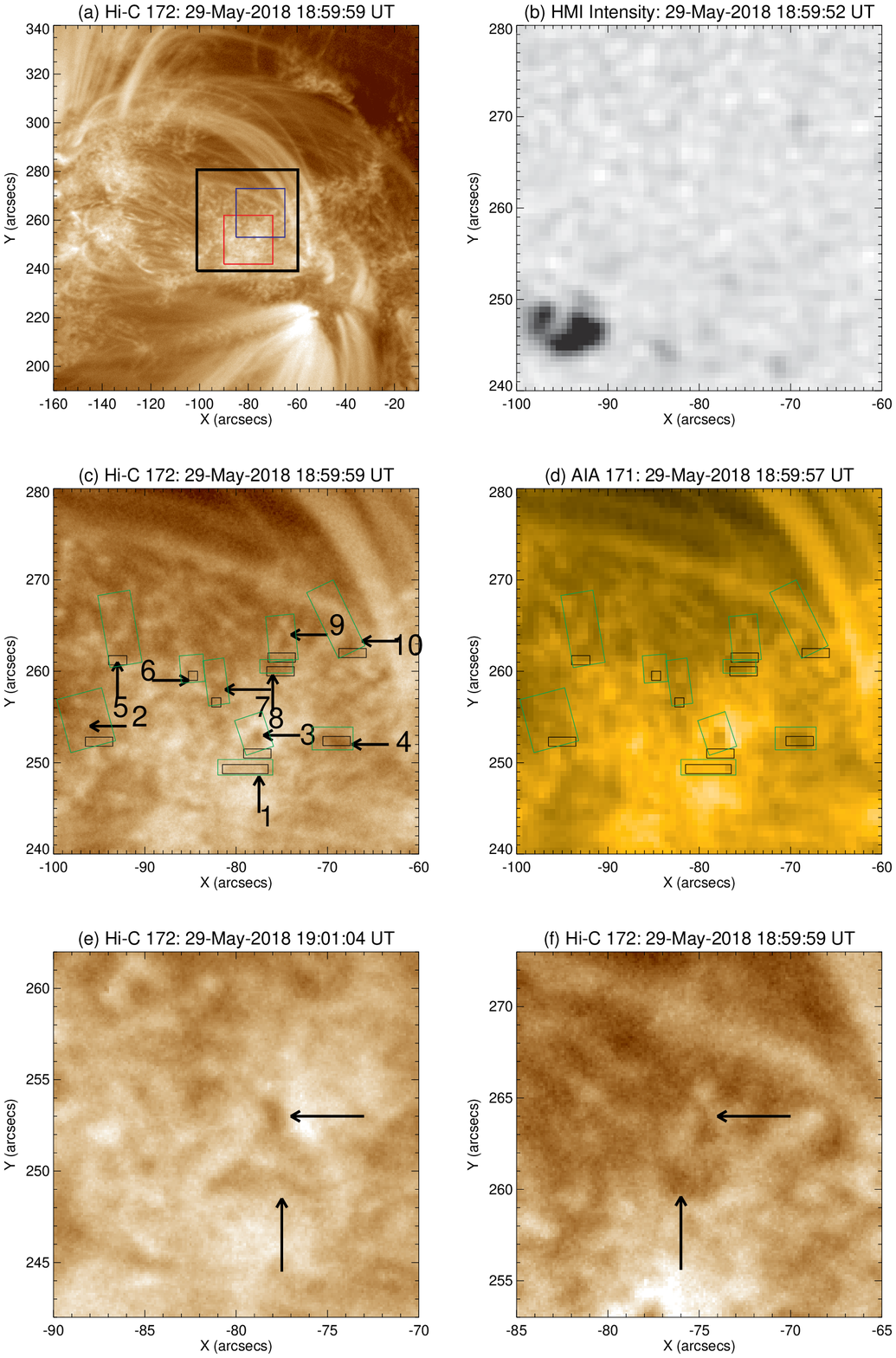}
\centerline{Figure~1}
\end{figure}
\clearpage

\begin{figure}
\epsscale{1.0}
\plotone{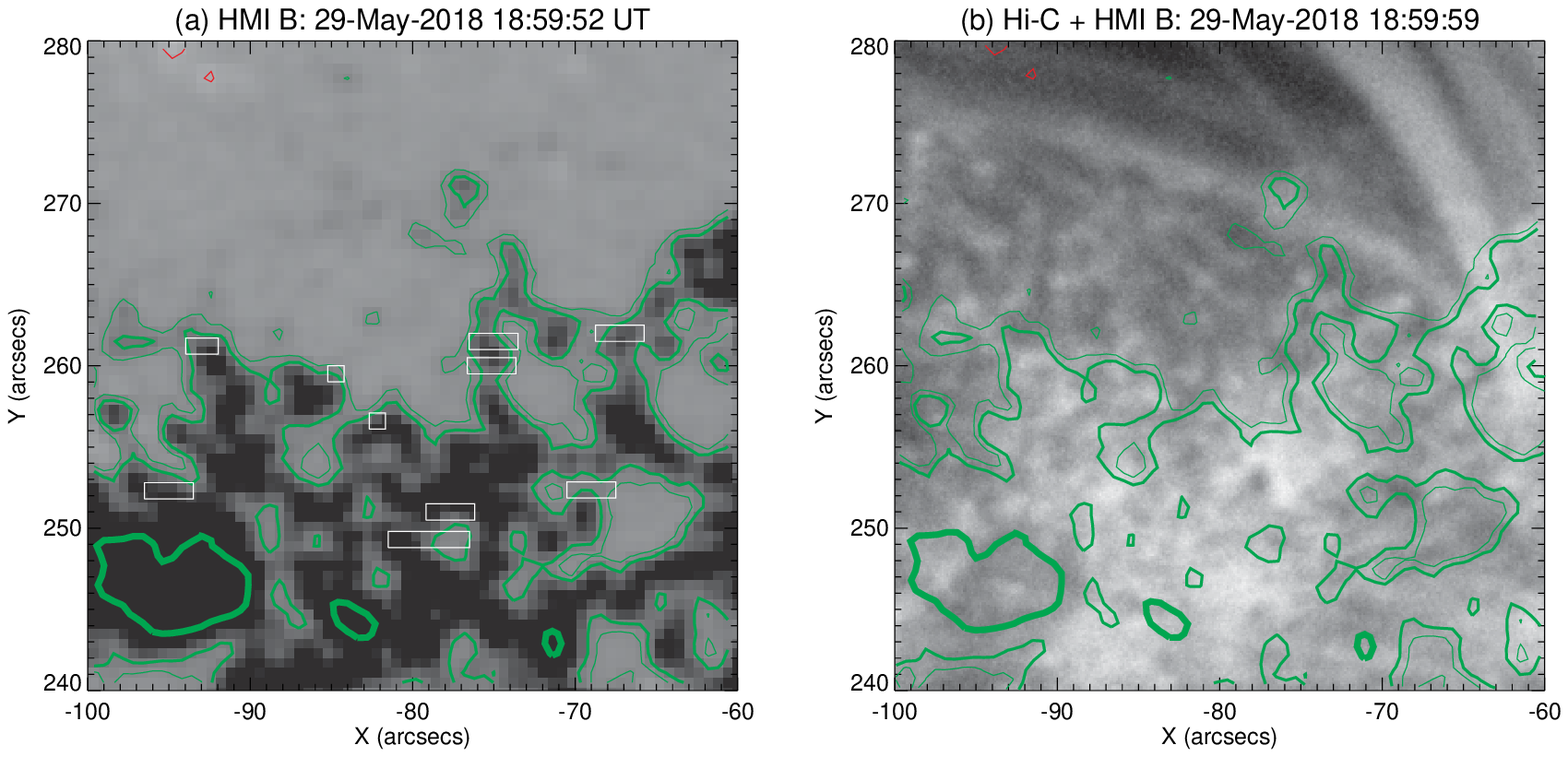}
\centerline{Figure~2}
\end{figure}
\clearpage

\begin{figure}
\epsscale{1.1}
\plotone{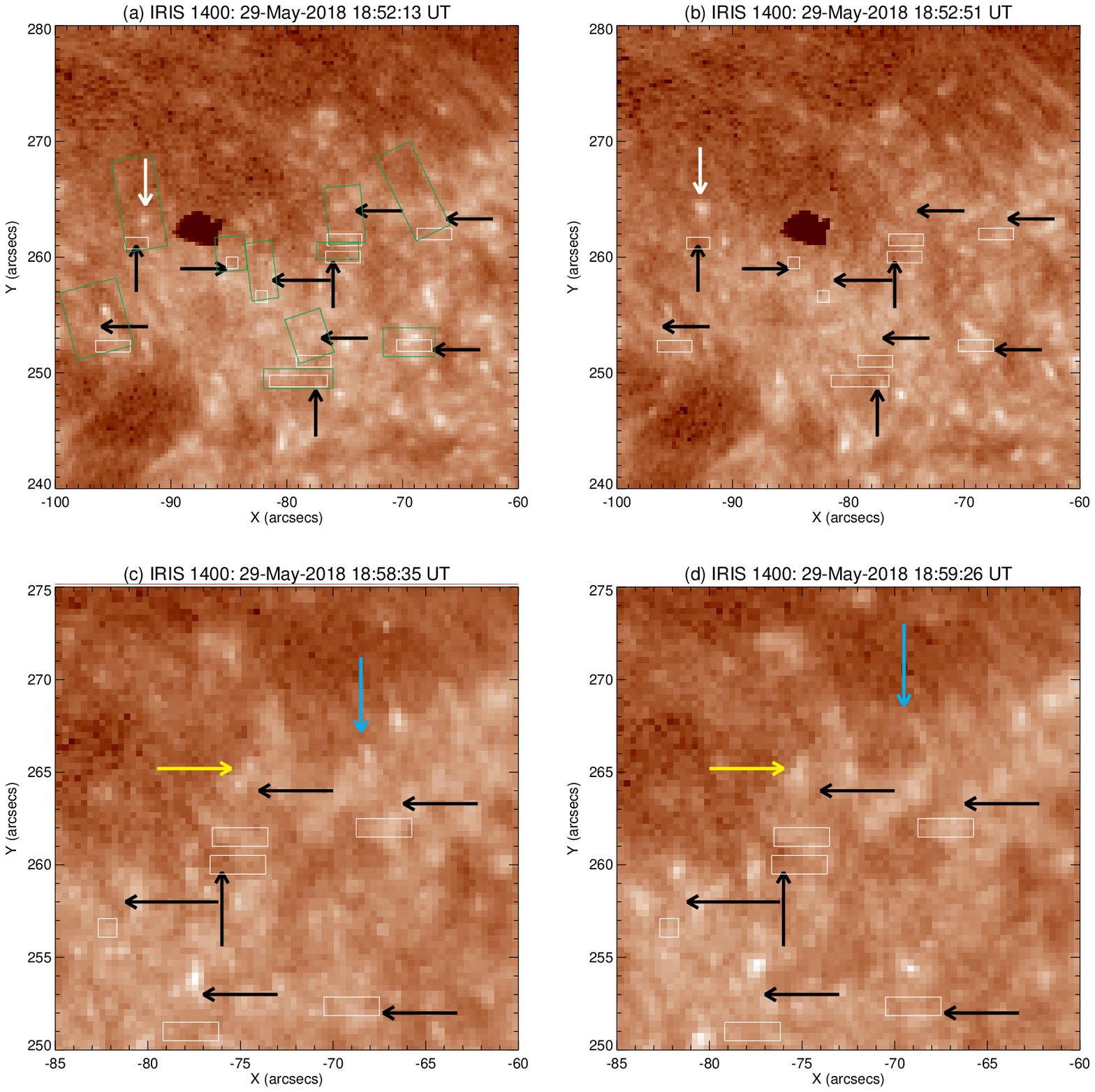}
\centerline{Figure~3}
\end{figure}
\clearpage

\begin{figure}
\epsscale{1.0}
\plotone{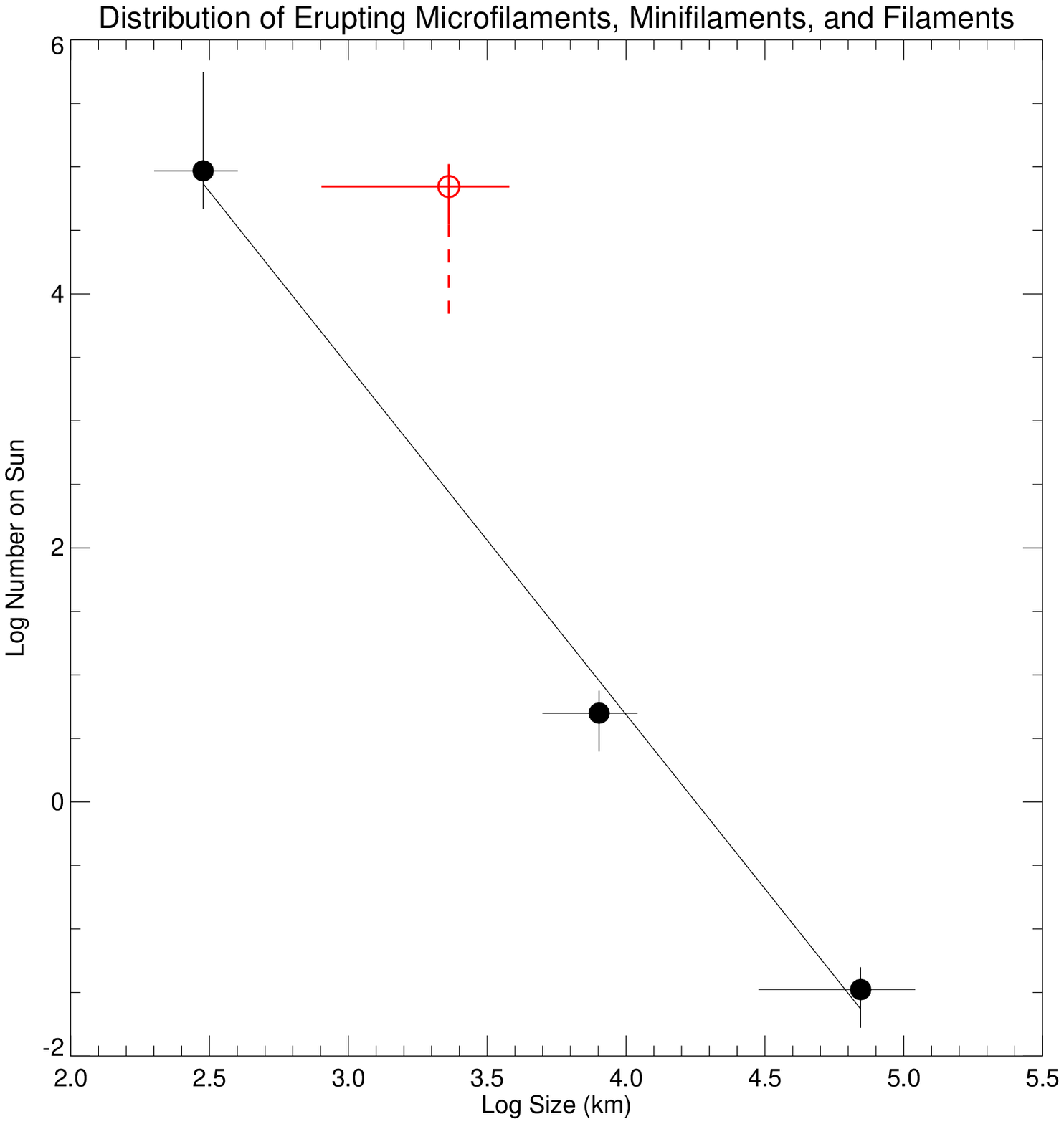}
\centerline{Figure~4}
\end{figure}
\clearpage

\end{document}